\documentstyle[twocolumn,seceq,epsbox]{jpsj}
\input epsf.tex

\def\runtitle{Non-Fermi Liquid Behavior of U-impurity with $f^2$-Singlet Ground State}
\def\runauthor{Satoshi {\sc Yotsuhashi}, Kazumasa {\sc Miyake} and Hiroaki {\sc Kusunose}}

\title{Non-Fermi Liquid Behavior of U-impurity with $f^2$-Singlet Ground State}

\author
{Satoshi {\sc Yotsuhashi}\footnote{E-mail: yotsu@eagle.mp.es.osaka-u.ac.jp}, Kazumasa {\sc Miyake}, and Hiroaki {\sc Kusunose}$^{1}$}
\inst
{Division of Materials Physics, Department of Physical Science, Graduated School of Engineering Science,\\ Osaka University, Toyonaka, Osaka 560-8531\\
$^1$Department of Physics, Tohoku University, Sendai, 980-8578}

\recdate{\today}

\abst
{It is shown by the Wilson's numerical renormalization-group approach that Non-Fermi liquid (NFL) behaviors arise even for U-impurity with an $f^2$-singlet ground state, and all the puzzling behaviors of R$_{1-x}$U$_x$Ru$_2$Si$_2$ (R=Th, Y and La, $x\le 0.07$) can be explained on its basis.
There exists an unstable NFL fixed point due to the balance between Kondo singlet and Crystalline-Electric-Field singlet ground states. For wide region in parameter space the system shows the NFL behavior as a transient phenomenon in temperature range accessible by experiment.
Results for temperature dependence of the resistivity, the magnetic susceptibility and the Sommerfeld coefficient with and without magnetic field reproduce the behaviors observed in Th$_{1-x}$U$_x$Ru$_2$Si$_2$.
Less anomalous properties of R$_{1-x}$U$_x$Ru$_2$Si$_2$ (R=Y and La) can be understood along the same scenario provided that those compounds have different sets of parameters.
}

\kword
{Non-Fermi liquid, numerical renormalization-group method,
 Crystalline-Electric-Field singlet,
 R$_{1-x}$U$_x$Ru$_2$Si$_2$ (R=Th, Y and La)}

\begin{document}
\sloppy
\maketitle
%
%
Recently, issues about the Non-Fermi liquid (NFL) behaviors in dilute U arroys have attracted much attention.
They are related to a general interest of Kondo phenomena \cite{Cox} which arise in some cases due to the existence of plural number of strongly correlated 5f electrons per U ion.
Especially, the behaviors in series of compounds R$_{1-x}$U$_x$Ru$_2$Si$_2$ (R=Th, Y and La, $x\le 0.07$) have been well investigated as a single-site effect, because anomalous properties of these materials are well scaled by impurity concentrations \cite{Ami94,Maru96,Ami00}.
The latter fact makes the theoretical approach simple compared with the system in which the intersite effect is considered to play some role, such as Y$_{1-x}$U$_x$Pd$_3$\cite{Sea91,And91}.
Th$_{1-x}$U$_x$Ru$_2$Si$_2$ shows $-\ln T$ divergence both of the susceptibility $\chi$ and the Sommerfeld coefficient $\gamma$ with decreasing temperature.
However, a big puzzle still remains as discussed shortly.

The logarithmic dependence of these thermodynamic quantities itself is consistent with the prediction based on the quadrupolar Kondo model \cite{Cox87}, if it is generalized appropriately \cite{Cox96,Kusu96}.
The quadrupolar Kondo effect is expected to arise for a non-Kramers doublet ground state of $f^2$-configuration and to be mapped onto the S=1/2 two-channel Kondo model \cite{Noz80}.
The exact solution of S=1/2 two-channel Kondo model predicts the asymptotic behavior of $\gamma$, $\chi$ $\propto -\ln T$ and the resistivity $\rho$ $\propto {\rm const}\pm \sqrt{T}$ in the limit $T \to 0$ with the residual entropy $S(T \to 0)=\frac{1}{2} \ln 2$ \cite{And84,Sac91,Aff93}.
Since R$_{1-x}$U$_x$Ru$_2$Si$_2$ (R=Th, Y and La) has the tetragonal symmetry and the valence of the uranium ion is expected to be mainly U$^{4+}$ (5$f^2$), many theoretical studies of this problem have been done on the basis of the two-channel Kondo model assuming the magnetic doublet ground state, $|\Gamma_5^{(2)}\pm \rangle$, in (5$f^2$) configuration \cite{Cox,Cox96,Sakai96,Suzu97,Shimi98,Shimi99}.

However, there exist puzzling properties about Th$_{1-x}$U$_x$Ru$_2$Si$_2$ which cannot be explained by the two-channel Kondo model.
First of all, $\chi$ shows the Curie law while $\rho$ shows large value of the order of the unitarity limit for $10$K$<T<100$K.
Namely, the susceptibility indicates the existence of the localized moment while the resistivity suggests the confinement of the localized moment by the spin polarization of conduction electrons.
This apparent inconsistency was removed by the extended two-channel Anderson model introduced and discussed by Sakai and his collaborators \cite{Sakai96,Suzu97,Shimi98}, who showed that there exists wide set of parameters which gives results consistent with the experiments of Th$_{1-x}$U$_x$Ru$_2$Si$_2$.
However, in the case of R$_{1-x}$U$_x$Ru$_2$Si$_2$ (R=Y and La), $\rho$ is almost constant below 10K, which cannot be solved by such an extended two-channel Anderson model.
Secondly, the divergence of $\gamma$ is suppressed for applied magnetic field $H$ of the order of several Tesla in Th$_{1-x}$U$_x$Ru$_2$Si$_2$ \cite{Ami00}.
In any theories based on the two-channel Kondo model, the entropy is expected to be released at finite temperature because the magnetic field lifts the degeneracy of the $|\Gamma_5^{(2)} \pm \rangle$ doublet ground state, leading enhancement of $\gamma$ \cite{Shimi99}.
Thus, the theory based on the two-channel Kondo model with magnetic doublet configuration does not seem to be a final theory for this problem.

The purpose of this paper is to solve this puzzle on the basis of a novel mechanism assuming Crystalline-Electric-Field (CEF) singlet ground state, in which the competition between the conventional Kondo singlet and CEF singlet gives rise to the NFL behavior in low but intermediate temperature regime as a transient phenomenon.
A system exhibiting such competition can be mapped to the $f^2$-impurity model with anisotropic antiferromagnetic Hund's-rule coupling and different hybridizations with conduction electrons, when CEF levels are specified by pseudo spins.
It has been recognized by recent studies \cite{Jones88,Jones89,Sakai92,Aff92,Aff95,yotsu01} that such models exhibit the NFL behavior reminiscent of two-channel Kondo model for certain critical set of parameters.
On the basis of Wilson's numerical renormalization-group (NRG) method \cite{Wilson75,Krish80,Sakai89}, we show that the NFL behaviors arise, as a transient phenomenon near the unstable fixed point, rather robustly even for parameters off the criticality in rather wide temperature region accessible by experiments.
Series of present results are consistent with those observed in R$_{1-x}$U$_x$Ru$_2$Si$_2$ (R=Th, Y and La).
Especially, all the anomalous properties of Th$_{1-x}$U$_x$Ru$_2$Si$_2$ can be explained semiquantitatively for a certain set of parameters.

%
%
We consider two low-lying $f^1$ doublet states out of three doublet of $j=5/2$ orbitals in the tetragonal symmetry.
These four (two doublet) states can be represented by pseudo spin:
\begin{eqnarray}
&&|\Gamma_7^{(2)}, +\rangle =-\sqrt{\frac{1}{6}}\left| \frac{5}{2}\right\rangle +\sqrt{\frac{5}{6}}\left| -\frac{3}{2}\right\rangle \equiv |\uparrow , 0 \rangle , \\
&&|\Gamma_7^{(2)}, -\rangle =\sqrt{\frac{1}{6}}\left| -\frac{5}{2}\right\rangle -\sqrt{\frac{5}{6}}\left| \frac{3}{2}\right\rangle \equiv |\downarrow, 0 \rangle , \\
&&|\Gamma_6, +\rangle = \left|+\frac{1}{2} \right\rangle \equiv |0, \uparrow \rangle, \\
&&|\Gamma_6, -\rangle = \left|-\frac{1}{2} \right\rangle \equiv |0, \downarrow \rangle,
\end{eqnarray}
where $|\uparrow, 0 \rangle$ ($|0, \uparrow \rangle$) represents the state that the pseudo spin in channel 1 (2) is up and the channel 2 (1) is empty, and notations are conventional otherwise.
In terms of these four states, the relevant CEF states in $f^2$-configuration are written in the $j$-$j$ coupling scheme as follows:
\begin{eqnarray}
&&|\Gamma_4 \rangle =\frac{1}{\sqrt{2}}\left(|2\rangle - |-2\rangle\right)=\frac{1}{\sqrt{2}}\left(|\uparrow, \downarrow \rangle - |\downarrow, \uparrow \rangle \right) \\
&&|\Gamma_3 \rangle =\frac{1}{\sqrt{2}}\left(|2\rangle + |-2\rangle\right)=\frac{1}{\sqrt{2}}\left(|\uparrow, \downarrow \rangle + |\downarrow, \uparrow \rangle \right) \\
&&|\Gamma_5^{(2)}, +\rangle =\beta | 3\rangle -\alpha | -1\rangle = |\uparrow , \uparrow \rangle , \\
&&|\Gamma_5^{(2)}, -\rangle =\beta |-3\rangle -\alpha | 1\rangle = |\downarrow , \downarrow \rangle .
\end{eqnarray}
Here, the energy levels of $|\uparrow\downarrow, 0 \rangle$ and $|0, \uparrow\downarrow \rangle$ have been omitted because of large intra-orbital Coulomb repulsion.
One of fundamental assumptions is to adopt CEF level scheme shown in Fig.1.
This level scheme can be reproduced by the pseudo-spin Hamiltonian
\begin{eqnarray}
H_{\rm Hund}=\frac{J_{\bot }}{2}[S_1^+ S_2^- +S_1^- S_2^+]+ J_z S_1^z S_2^z ,
\label{HHund}
\end{eqnarray}
where $\vec{S}_m$ denotes a pseudo-spin operator of the localized electrons in the orbital $m$.
By taking such the pseudo spin representation, 
it can be understood clearly that 
the CEF singlet competes with the Kondo singlet directly and 
we can perform calculations more accurately owing to an appearance of 
new conserved quantity, pseudo spin.
The energy separation between the singlet ($\Gamma_4$) and the doublet ($\Gamma_5^{(2)} \pm$) is set to be $\Delta$ and that between the $\Gamma_4$ and the singlet ($\Gamma_3$) is set to be $K$.
The couplings $J_{\bot}$ and $J_z$ are related to $(K,\Delta)$ as $J_{\bot }$=$K$ and $J_z$=$2\Delta -K$.
It is remarked that (\ref{HHund}) represents an anisotropic antiferromagnetic Hund's-rule coupling, which cannot exist in real spin system.

With this preliminary, the system in question can be described by a conventional two-channel Anderson model with {\it antiferromagnetic} Hund's-rule coupling:
\begin{eqnarray}
&&H=H_{\rm K}+H_{\rm mix}+H_{\rm f}+H_{\rm Hund}, \label{H}
\end{eqnarray}
with
\begin{eqnarray}
&&H_{\rm K}\equiv \sum_{m=1,2} \sum_{\vec{k} \sigma}\epsilon_{\vec{k}} c_{\vec{k}m\sigma}^\dagger c_{\vec{k}m\sigma}, \\
&&H_{\rm mix}\equiv \sum_{m=1,2} \sum_{\vec{k}\sigma}(V_{\vec{k},m} c_{\vec{k}m\sigma}^\dagger f_{m\sigma}+{\rm h.c}.), \\
&&H_{\rm f}\equiv \sum_{m\sigma} E_{{\rm f}m} f_{m\sigma}^\dagger f_{m\sigma}+\sum_{m \sigma} \frac{U_{m}}{2} f_{m\sigma}^\dagger f_{m\bar{\sigma}}^\dagger f_{m\bar{\sigma}} f_{m\sigma}.
\end{eqnarray}
It is noted that the hybridization $V_{{\vec{k},m}}$ depends only on its channel index $m (=1,2)$ and the inter-orbital Coulomb interaction is neglected for simplicity.
Hereafter, we take the unit of energy as $(1+\Lambda^{-1}) D / 2$ where $D$ is half the bandwidth of conduction electrons and $\Lambda$ is a discretization parameter in NRG calculation.
In the case $K$$=$$\Delta$$=$$0$, this model reduces to two independent 
conventional Anderson models.
Therefore, we can restrict ourselves to the case that each $f$-orbital is almost singly occupied by taking parameters such that  $E_{\rm f1}=E_{\rm f2}=-0.4$, $U_1=U_2=2.0$, $V_1=0.7$, and $V_2=0.34$.

The system described by (\ref{H}) shows the competition between the Kondo singlet and the local singlet \cite{yotsu01,Kura92} as seen in two-impurity systems \cite{Jones88,Jones89,Sakai92,Aff92,Aff95}.
Two types of ground state are possible, i.e., (i) the Kondo singlet state characterized by total phase shift being equal to $\pi$ (i.e., $\delta_1=\delta_2=\pi /2$), and (ii) the local singlet state characterized by $\delta_1=\delta_2=0$.
Then, there exists a locus of unstable fixed points characterized by $\delta_1=\delta_2=\pi/4$, and on which the NFL behavior appears.
A critical line can be determined by the NRG calculation, across which the even-odd alternation in energy spectrum is interchanged.
For the present parameter set, the critical line is shown in Fig. 2.
When initial couplings $(K,\Delta)$ are located near the critical line, the renormalization flow passes near the NFL unstable fixed point, and then arrives at the Fermi liquid (FL) fixed point.
Namely, the NFL behavior can be seen in the observable temperature range, if $(K,\Delta)$ are close enough to the critical line.
Although we present here the results for only one parameter set, 
we have confirmed that the NFL behavior can be seen rather robustly 
at finite temperature even if the conditions, such as 
particle-hole symmetry, spin rotational symmetry, and 
identical $T_{\rm K}$'s in each orbital, are not satisfied.

Another fundamental assumption of the present theory is that Kondo temperatures of two channels are well separated as $T_{\rm K1}\gg T_{\rm K2}$ due to the difference of hybridization ($T_{\rm K1}=1.50\times 10^{-1}$ and $T_{\rm K2}=8.27\times 10^{-4}$ for $J_\perp=J_z=0$).
Thus, below $T_{\rm K1}$, $\vec{S}_1$ has been screened out by conduction electrons in channel 1, while $\vec{S}_2$ is still active as the localized moment.
As the temperature decreases further, the system sets in the critical region, where various quantities exhibit the NFL behaviors.
The characteristic energy $T_{\rm K}$ is given by that of the S=1/2 two-channel Kondo model with the exchange coupling $J\sim |V_2|^2/\max(E_{f2},U_2)$\cite{Pang91,Aff92t,yotsu01,yotsu02} because the NFL behavior arises from the balance of two couplings by which $\vec{S}_2$ interacts with the two ``conduction'' electron channels\cite{yotsu01,yotsu02}.
Below the crossover temperature $T_{\rm cr}$ at which $\chi$ saturates and $\tau^{-1}(\omega)$ ceases to change simultaneously (as shown later in Fig.4), the system flows into the local singlet ground state.
Namely, $T_{\rm cr}$ measures the ``distance'' from the critical point.
In fact, as the parameter set $(K,\Delta)$ goes away from the critical boundary, $T_{\rm cr}$ increases.
We have checked this by series of numerical calculations.
Note that the parameter set locating opposite side of the critical boundary leads to the Kondo singlet ground state.
In the present case, it is estimated that $T_{\rm K}\sim 10^{-3}$ and 
$T_{\rm cr}\sim 10^{-5}$, which is assumed to be the case of 
Th$_{1-x}$U$_x$Ru$_2$Si$_2$.
The renormalization flow in terms of the phase shift is schematically shown in Fig.3.

%
%
We show the relation between the magnetic susceptibility, $\chi(T)$, and the total scattering rate at $T=0$, $\tau^{-1}(\omega)$, for $K=0.14$ and $\Delta=0.08$ in Fig.4.
The temperature dependence of the resistivity may infer from the $\tau^{-1}(\omega)$, which is calculated by sum of each channel contributions $\tau_m^{-1}=|V_m|^2A_m(\omega)$, $A_m(\omega)$ being the single-particle spectral function.
The effect of magnetic field has been taken into account through the Zeeman coupling $-(g\mu_{\rm B})j_z H_z$ of each $f$-orbital with $j=5/2$ and $g=6/7$.
For $T_K<(T, \omega)<T_{K1}$, $\chi(T)$ shows the Curie-law behavior, while $\tau^{-1}(\omega)$ takes a broad maximum of the order of the unitarity limit.
Below $T_K$, $\chi(T)$ increases logarithmically and $\tau^{-1}(\omega)$ decreases.
The crossover to the local singlet ground state has not been observed in Th$_{1-x}$U$_x$Ru$_2$Si$_2$, which suggests that $T_{\rm cr}$ is well below the lowest temperature.
Behaviors shown in Fig.4 well simulates those observed in 
R$_{1-x}$U$_x$Ru$_2$Si$_2$ (R=Th, Y and La) \cite{Ami00}.
It is emphasized that the adopted $K=0.14$ and $\Delta =0.08$ is not located so close to the critical line.

The contributions of each channels $\tau_m^{-1}(\omega)$ to $\tau^{-1}(\omega)$ is shown in Fig.5.
The rather flat bump of $\tau^{-1}(\omega)$ in the region $\omega<T_{K1}$ arises from the cancellation of decreasing $\tau^{-1}_1(\omega)$ and increasing $\tau^{-1}_2(\omega)$.
The gradual decrease of $\tau_1^{-1}$ can be understood by an exchange field brought about by growing renormalized Hund's-rule coupling, which destroys the Kondo singlet in channel 1.
The further flat bottom of $\tau^{-1}(\omega)$ below $T_{\rm cr}$ 
indicates that the system flows to the local singlet ground state.

The temperature dependence of the Sommerfeld coefficient $\gamma$$\equiv$$C/T$ and the entropy under the magnetic field is shown in Fig.6.
By applying the magnetic field of $10^{-3}D\sim$ several Tesla, $\gamma$ is suppressed at low temperatures $T<T_{\rm K}$ because $\gamma$ shows stronger 
divergence than $-\ln T$ in the case of $H=0$.
This divergence is caused by a transient phenomenon from 
a NFL unstable fixed point to the FL fixed point.
The change of the entropy due to the magnetic field is moderate compared with that of the two-channel Kondo model with magnetic doublet.
This is because the magnetic field has little influence on the CEF singlet ground state.
Namely, the behaviors of the present model are relatively robust against the magnetic field.

%
%
In summary, we have shown that the puzzling problem of Th$_{1-x}$U$_x$Ru$_2$Si$_2$ ($x\le 0.07$) can be fully understood on the basis of a mechanism characterized by Non-Fermi liquid like transient phenomenon with $f^2$-singlet ground state.
A key assumption is that the system corresponding to Th$_{1-x}$U$_x$Ru$_2$Si$_2$ is incidentally located near the critical points.
However, it is not so artificial to assume such an incident if one considers the fact that the NFL behaviors appear weakly in much more restricted temperature range for R$_{1-x}$U$_x$Ru$_2$Si$_2$ (R=Y and La) which would have different set of parameters due to the difference of conduction band structure.
Experimental results show that $T_{\rm cr}$ in the case of R=La is about $10$K while that in the case of R=Th has not been observed.
This indicates that the parameter set of La$_{1-x}$U$_x$Ru$_2$Si$_2$ would be located less close to the critical points than that of Th$_{1-x}$U$_x$Ru$_2$Si$_2$.
If $T_{\rm K}$'s are increased with unchanged CEF level structure,
the final fixed point can be changed from the CEF singlet to the Kondo singlet.
Experimentally, when high pressure is applied, these materials may show the increase of the resistivity with decreasing temperature
corresponding to the change of the fixed point.

%
%
We would like to thank O. Sakai and H. Amitsuka for variable comments.
One of us (S.Y.) acknowledges useful conversations with H. Maebashi.
This work is supported by a Grant-in-Aid for COE Research (No.10CE2004) of the Ministry of Education, Science, Sports, Culture and Technology.

\begin{figure}[htbp]
\begin{center}
\epsfxsize=7.5cm \epsfbox{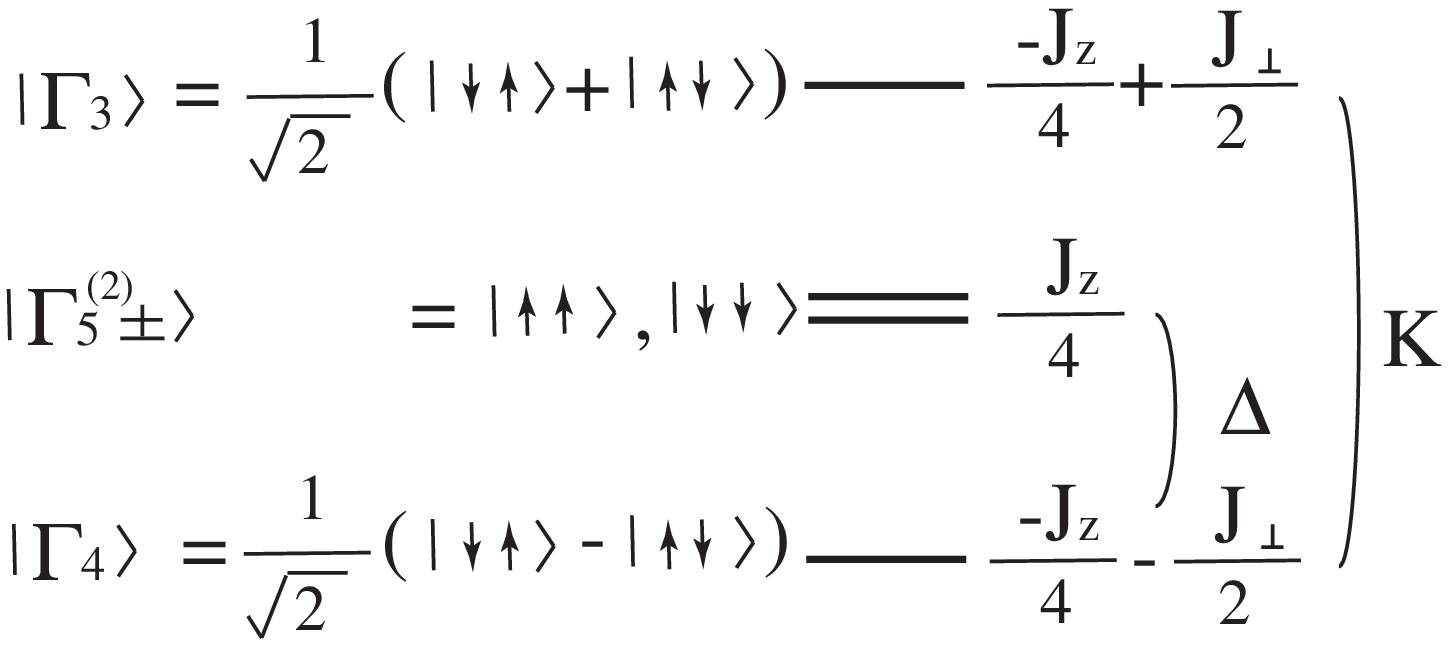}
\end{center}
	\caption{CEF level scheme in $f^2$-configuration adopted in this paper.} \label{fig1}
\end{figure}
\begin{figure}[htbp]
\begin{center}
\epsfxsize=7.5cm \epsfbox{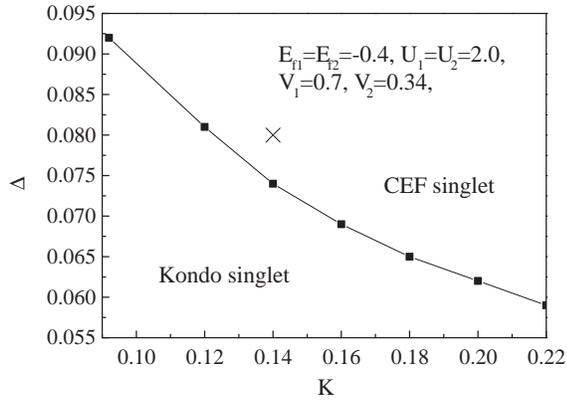}
\end{center}
	\caption{Phase diagram of the ground state in $K$-$\Delta$ plane. Squares represent the critical points determined by NRG calculations. The cross represents a parameter set used in this paper.} \label{fig2}
\end{figure}
\begin{figure}[htbp]
\begin{center}
\epsfxsize=5.5cm \epsfbox{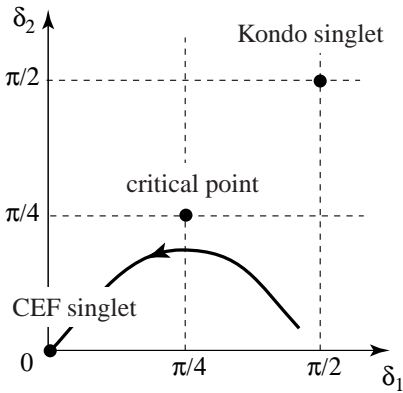}
\end{center}
	\caption{The renormalization flow below $T<T_{K1}$. The NFL behavior should be observed when the system passes near the critical point.} \label{fig3}
\end{figure}
\begin{figure}[htbp]
\begin{center}
\epsfxsize=7.5cm \epsfbox{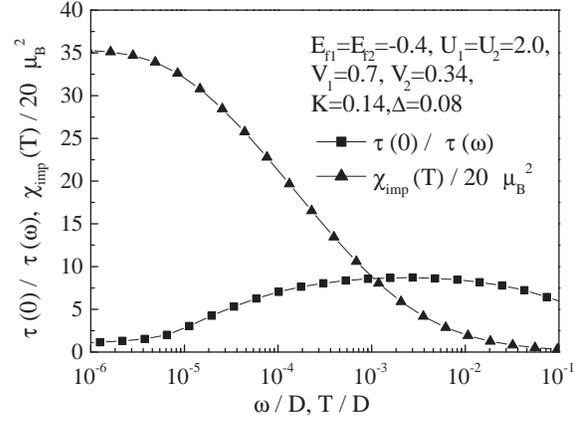}
\end{center}
	\caption{Temperature dependence of the magnetic susceptibility $\chi(T)$ (triangles) and frequency dependence of the scattering rate $\tau^{-1} (\omega)$ at $T=0$ (squares).} \label{fig4}
\end{figure}
\begin{figure}[htbp]
\begin{center}
\epsfxsize=7.5cm \epsfbox{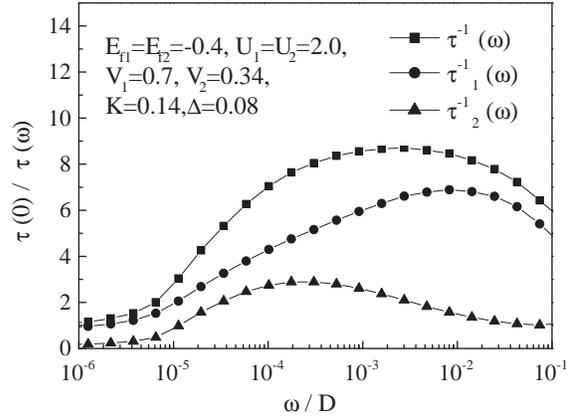}
\end{center}
	\caption{Contribution of each channels $\tau_1^{-1} (\omega)$ (circles) and $\tau_2^{-1} (\omega)$ (triangles) to the total scattering rate $\tau^{-1} (\omega)$ (squares).} \label{fig5}
\end{figure}
\begin{figure}[htbp]
\begin{center}
\epsfxsize=7.5cm \epsfbox{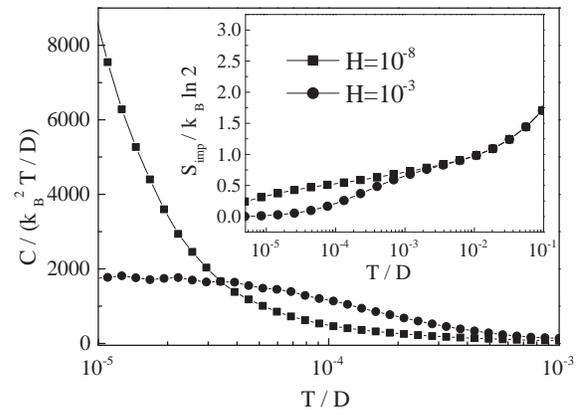}
\end{center}
	\caption{Temperature dependence of the Sommerfeld coefficient $\gamma \equiv C/T$ under the magnetic field of $H=10^{-8}$ (squares) and $H=10^{-3}$ (circles). The inset shows temperature dependence of the entropy. The other parameter set is the same as shown in Fig.4 and Fig.5.} \label{fig6}
\end{figure}

\end{document}